\documentclass{ws-ijbc}
\usepackage{ws-rotating}     

\usepackage[utf8x]{inputenc}
\usepackage{graphicx}
\usepackage{color}	

\begin{document}

\catchline{}{}{}{}{} 

\markboth{A. Navas {\em et. al}}{Functional Hubs in Mild Cognitive Impairment}

\title{Functional Hubs in Mild Cognitive Impairment\\}

\author{Adri\'an Navas${^*}$, David Papo, Stefano Boccaletti, F. del-Pozo}
\address{Centre for Biomedical Technology (UPM)\\
28922 Pozuelo de Alarc\'on, Madrid (Spain)\\
${^*}$adrian.navas@ctb.upm.es
}

\author{Ricardo Bajo, Fernando Maest\'u}
\address{Cognitive and Computational Neuroscience Lab\\
Centre for Biomedical Technology (UPM)\\
28922 Pozuelo de Alarc\'on, Madrid (Spain)\\
}

\author{Pedro Gil}
\address{
Memory Unit, Hospital Cl\'{\i}nico San Carlos\\ 
28040 Madrid, (Spain)}

\author{Irene Sendi\~na-Nadal and Javier M. Buld\'u}
\address{Complex Systems Group\\ 
URJC, 28933 M\'ostoles (Spain) and\\
Laboratory of Biological Networks\\
Centre for Biomedical Technology (UPM)\\
28922 Pozuelo de Alarc\'on, Madrid (Spain)\\}

\maketitle

\begin{history}
\received{(to be inserted by publisher)}
\end{history}

\begin{abstract}

We investigate how hubs of functional brain networks are modified as a result of mild cognitive impairment (MCI), a condition causing a
slight but noticeable decline in cognitive abilities, which sometimes precedes the onset of Alzheimer's disease. We used
magnetoencephalography (MEG) to investigate the functional brain networks of a group of patients suffering from MCI and a control group of
healthy subjects, during the execution of a short-term memory task. Couplings between brain sites were evaluated using synchronization
likelihood, from which a network of functional interdependencies was constructed and the {\em centrality}, i.e. importance, of their nodes
quantified. The results showed that, with respect to healthy controls, MCI patients were associated with decreases and increases in hub
centrality respectively in occipital and central scalp regions, supporting the hypothesis that MCI modifies functional brain network
topology, leading to more random structures. 

\end{abstract}


\section{Introduction}

\noindent Characterizing how the brain organizes its activity to carry out complex cognitive tasks is a highly non trivial pursuit. While early neuroimaging studies typically aimed at identifying patches of task-specific activation or local time-varying patterns of activity, it is now recognized that both at rest and during behaviour, that brain activity is characterized by the formation and dissolution of functionally meaningful integrated activity \cite{bre2002}. 

Over the last ten years, functional activity has started been investigated using a statistical physics understanding of a very old branch of pure mathematics: graph theory. This approach has extensively been applied to complex physical and biological 
systems \cite{new2003,boc2006,alb2007}, and represents a shift from a parallel distributed, computer-like to a complex system vision of the brain. In this approach, functional activity is thought of as a network. A network consists of a number of units (nodes) interacting by links. The specific pattern of connections between the nodes defines the network’s topology.  Network theory provides suitable descriptors at all scales from local microscopic, to mesoscopic, to global macroscopic ones. Models to reproduce global topological properties of natural systems e.g. the “small-world” \cite{wat1998} and “scale-free” \cite{bar1999} properties, and measures to properly quantify these and other observed features have been proposed \cite{boc2006}. 
The existence of central regions in brain functional networks and the way of quantifying their importance in the network structure has been the object of numerous studies \cite{spo2007,ach2006,buc2009,zam2010,joy2010,loh2010,hwa2012,kuh2012}. The centrality (i.e. importance) of a node $i$ can {\it prima facie} be measured by the number of connections $K(i)$ (degree) in the functional network, indicating how many regions are coordinated with it during a certain task. Once the amount of synchrony is quantified, a weighted network of connections can be constructed. In this case, the node strength $S(i)$, i.e. the sum of the weights of its links, becomes a more accurate indicator of the importance of a node in the network. Nevertheless, both the degree and the strength are local measures, which do not take into account the global topology of the functional network. 
To overcome this issue, global measures of centrality have been proposed. Node closeness $C(i)$, takes into account the number of steps that we have to do to go from one to another: the lower number of steps, the higher closeness a node has. The eigenvector centrality $E(i)$, is another indicator of the global relevance of a node and it is measured by computing the eigenvector $v_1$ associated to the first eigenvalue of the connectivity matrix (i.e., the matrix of interactions between nodes). A more intuitive measure is represented 
by node betweenness $B(i)$, which quantifies the number of shortest paths (minimum number of steps between two nodes) that cross a certain node, thus reflecting its importance in the transmission of information in the whole network. Closeness, eigenvector centrality and betweenness, make use of information about the structure of the whole network, nevertheless node centrality could be restricted solely to a certain region or community. To measure the community importance of nodes, Guimer\`a {\em et al.} \cite{gui2005} introduced two indicators, the within-module-degree $z_i$ and the participation coefficient $p_i$. The combination of both parameters not only measures the importance of a node inside its community but allows classifying the role played by the central nodes (hubs) in the overall community structure.

The application of these measures to functional brain networks may help in quantifying how relevant nodes are impaired by the emergence of different brain diseases. Here, we analyze the alteration of functional networks' hubs caused by mild cognitive impairment (MCI), a brain syndrome involving cognitive impairments beyond those expected based on the age and education of the individual often representing a transitional stage between normal aging and dementia. 
The progressive accumulation of the beta amyloid protein and the loss of cells and synapses along the Alzheimer's disease (AD) continuum impair cognitive function as a reflection of network malfunctioning. Thus, although hubs organization has been studied for fully-fledged AD \cite{buc2009,sta2009,deh2012}, whether hub organization impairment at the early stages of the disease such as in MCI. 
The MEG signal of seventeen patients suffering from MCI and seventeen control patients, was recorded during the execution of a memory task (see \cite{bul2011} for details). Next, we measured the synchronization likelihood ($SL$) between all pairs of nodes, and obtained a weighted correlation matrix that was analyzed by using Complex Networks Theory methods. 
In Buld\'u {\em et al.}, the attention was devoted to the global properties of the functional networks. Nevertheless, how hub organization is affected by the disease was not considered. In the present work we focus on how the role of the hubs is distorted by MCI.
Our hub characterization relies on various parameters, but we show how the eigenvector centrality $E(i)$ of the nodes, a global measure which quantifies the importance of a node in the whole structure of the network, is the more adequate indicator, in contrast with local measures such as degree or strength centrality. 
Therefore, we use this measure of centrality to identify (and quantify) network hubs.
Eigenvector centrality allows detecting the dominant regions in the functional connectivity network and, next, how these regions are affected by the disease. Our results show that the leading role played by network hubs is attenuated by MCI, which transforms the network into a more homogeneous one. These results are consistent with previously reported analyses showing that MCI increases the randomness of the global structure of the functional network \cite{bul2011}, thus, leading to a more homogeneous connectivity network.

The manuscript is organized as follows: in Section \ref{functional} we explain the methodology followed to obtain functional networks from magnetoencephalography measurements. Section \ref{methods} includes all
information related with the experimental procedure. Section \ref{centrality} contains a detailed explanation of the different centrality measures used in this work together with the main results. Finally, in Section 
\ref{discussion}, we discuss the implications of the results obtained and the relation with other studies
of hub characterization in functional brain networks. 


\section{Obtaining functional networks}
\label{functional}

\begin{figure}[htbp]
\begin{center}
\vskip1.2cm
\psfig{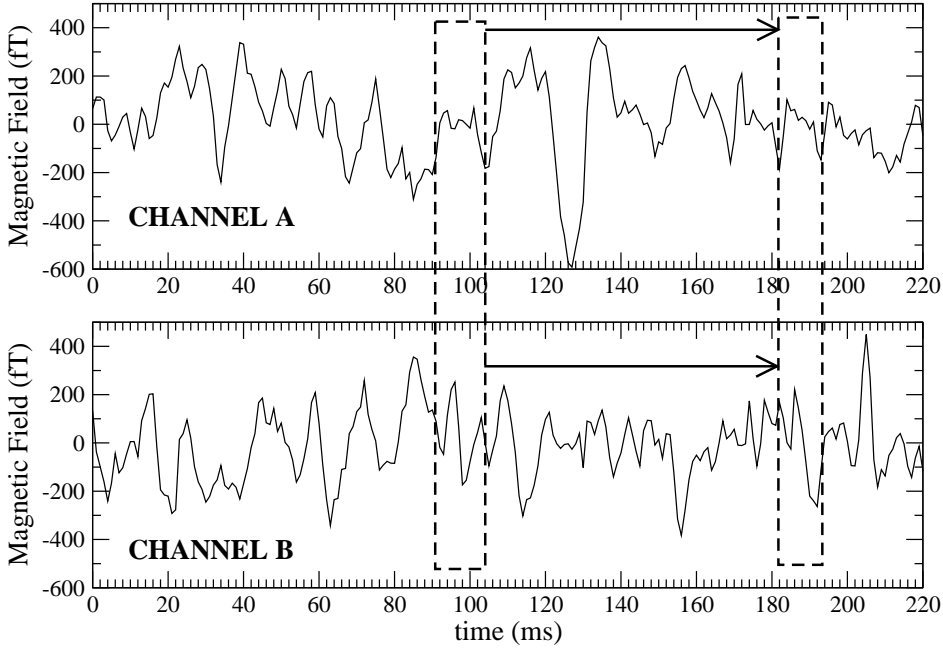}
\vskip0.5cm
\end{center}
\caption{Synchronization likelihood ($SL$) between two brain regions. Time series correspond to the recorded electromagnetic field
of two brain sites. The $SL$ algorithm detects windows of repeated patterns within the time series of channel A and, next, checks whether the
channel B also shows a repeated pattern at the same time windows, no matter if it is the same or different to that observed in channel A.
Values of $SL$ close to one (zero), indicate a high (low) correlation.}
\label{fig:fig01}
\end{figure}

Functional brain networks rely on the coordinated activity between brain regions, which requires to quantify the 
synchronization of each pair of regions, i.e. nodes, within the network. There exist a number of different
measures of synchronization \cite{boc2002,per2005} that can be applied to the recorded time series of each node (see Section \ref{methods} for details on how the dynamics of the nodes are obtained). Among them, we have 
chosen Synchronization Likelihood ($SL$) \cite{sta2002}, a nonlinear measure of the synchronized activity that has been proven to be a suitable quantifier for datasets obtained from 
magnetoencephalographic recordings \cite{sta2006,bul2011}. This index, which is closely related to the concept of generalized mutual information \cite{buz1994}, relies on the detection of simultaneously occurring patterns, which can be complex and widely different for
every pair of signals (see the example of Fig. \ref{fig:fig01}). Let \textbf{X} denote the matrix containing all the $M$ signals (one per each channel that measures a certain cortical region) of $n$ time steps, and let $X_n=[x_n, y_n,...,z_n]$, where $X_{1,n}=x_n$, $X_{2,n}=y_n$ and so on. In the time series of a given channel, for each time step $n$, we define the probability that embedded vectors are closer to each other than a distance $\varepsilon$: 

\begin{equation}
P_{n}^{\varepsilon}\left( X\right) =\dfrac {1} {2\left( w_{2}-w_{1}\right)} \sum_{\substack{m=1\\w_1<\vert n-m\vert<w_2}}^N\theta \left( \varepsilon -\left| x_{n}-x_{m}\right| \right) 
\end{equation}

where $\vert.\vert$ is the Euclidean distance and $\theta$ is the Heaviside step function, being $\theta(x)=0$ if $x\leq0$ and $\theta(x)=1$ for $x>0$. Here $w_1$ and $w_2$ are two windows: $w_1$ is the Theiler correction for autocorrelation effects and should be at least of the order of the autocorrelation time; $w_2$ is a window that sharpens the time resolution of the synchronization measure and is chosen such that $w_1\ll w_2\ll N$.

Now, for each signal and each time $n$, the critical distance $\varepsilon_n$ is determined for which $P_{n}^{\varepsilon}(x)= p_{ref}$, where $p_{ref}\ll1$. We can determine for each discrete time pair $(n,m)$ within our considered window $(w_1 < \vert n-m \vert <w_2)$ the number of channels $H_{n,m}$ where the embedded vectors $x_{k,n}$ and $x_{k,m}$ will be closer together than this critical distance $\varepsilon_{k,i}$: 

\begin{equation}
H_{n,m}=\sum _{k=1}^{M}\theta \left( \varepsilon _{k,n}-\left| x_{k,n}-x_{k,m}\right| \right)
\end{equation}

This number lies in a range between $0$ and $M$, and reflects how many of the embedded signals ``resemble'' each other.

We can now define a \textit{synchronization likelihood} $SL_{n,m}(X)$ for each channel $k$ and each discrete time pair $(n,m)$ as:

\begin{equation}
\begin{aligned}
& \mathrm{if} \left| x_{n}-x_{m}\right| < \varepsilon_{n}:SL_{n,m}\left( X\right) =\dfrac {H_{n,m}-1} {M-1}\\
& \mathrm{if} \left| x_{n}-x_{m}\right| \geq \varepsilon_{n}:SL_{n,m}\left( X\right) =0
\end{aligned}
\end{equation}

By averaging over all $m$, we finally obtain the synchronization likelihood, $SL_n(X)$: 

\begin{equation}
SL_{n}\left(X\right) =\dfrac {1} {2\left( w_{2}-w_{1}\right)} \sum_{\substack{m=1\\w_1<\vert n-m\vert<w_2}}^N SL_{n,m} \left( X \right) 
\end{equation}

Synchronization likelihood $SL_n(X)$ describes how strongly channel $x$ at time $n$ is synchronized to all the other $ M-1 $ channels.
The range of values of $SL$ is $0\leq SL\leq 1$, being ($p_{ref}$) when all $M$ time series are uncorrelated, and 1 for maximal synchronization of all $M$ time series. The value of $p_{ref}$ can be set at an arbitrarily low level, and does not depend on the properties of the time series, nor it is influenced by the embedding parameters \cite{sta2002}.

\begin{figure}[htbp]
\begin{center}
\psfig{file=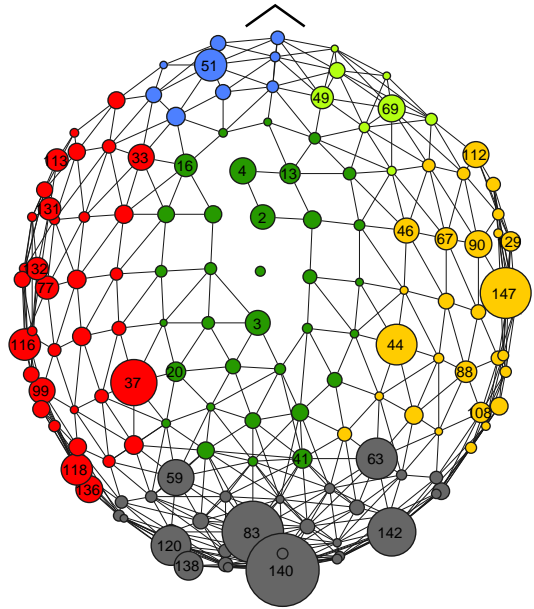,width=0.4\textwidth}
\end{center}
\caption{Averaged functional network of the control group. Only $5\%$ of the links with higher weights have been plotted. Colors indicate the 
lobe that a node belongs to: frontal-left (blue), frontal-right (light green), central (dark green), temporal left (red), temporal-right
(light orange) and occipital (dark grey). Node size is proportional to the node strength S(i). Note the higher density of connections
at the occipital and frontal lobes. Node number is indicated for those nodes with higher strength.}
\label{fig:fig02}
\end{figure}

The $SL$ yields a symmetric and weighted correlation matrix $w_{ij}$, which can be analyzed using Complex Networks metrics \cite{boc2006}. 
Before computing their network parameters, we follow the normalization technique proposed in \cite{bul2011} in order to avoid intrinsic differences from different individuals. The off-diagonal weights in the correlation matrices are rescaled to the interval $[0,1]$ by means of

\begin{equation}
SL_{ij}=\dfrac{w_{ij}-min(w_{ij})}{ max(w_{ij})-min(w_{ij})}
\end{equation}

\noindent while the diagonal is set to zero after normalization. 
Figure \ref{fig:fig02} shows an example of a functional brain network obtained from a control group of seventeen healthy individuals
(see Section \ref{methods} for details).
The normalized weighted correlation matrix $W_{ij}$ leads to a fully connected network (since all pair of nodes have a $SL$ $\geq 0$), so only
the $5\%$ of the links with higher weights (i.e., $SL$) have been plotted to ease visualization of the network. The size of a node $i$ is proportional to its strength $S(i)$, defined as the sum of the weights of all its connections.
We can observe how the density of connections is higher at the peripheral regions and, specially, at the occipital lobe, while the central
cortical region is sparsely connected (note that we are only considering the links with higher correlations). Interestingly, nodes with higher strengths,
i.e., the network hubs, are mainly localized in the occipital lobe.

\section{Materials and Methods}
\label{methods}
 
The magnetoencephalographic (MEG) signal was recorded with a 256 $Hz$ sampling frequency and a band pass of [$0.5,50$] $Hz$, using a $148$-channel whole-head magnetometer (MAGNES\copyright $2500$ WH, $4-D$ Neuroimaging) confined in a magnetically shielded room. Seventeen MCI patients and seventeen healthy volunteers were recorded during a Sternberg's letter-probe task \cite{det1991,mae2001}.

MCI diagnosis was established according to the criteria proposed by Petersen {\it et al.} \cite{gru2004,pet2004}. Thus, MCI patients fulfilled the following criteria: (1) cognitive complaint corroborated by an informant (a person who stays with the patient at least for half a day at least 4 days a week); (2) objective cognitive impairment, documented by delayed recall in the logical memory II sub-test of the revised Wechsler Memory Scale (score $\leq16/50$ for patients with more than 15 years of education; score $\leq8/50$ for patients with $8-15$ years of education); (3) normal general cognitive function, as assessed by a clinician during a structured interview with the patient and an informant and, additionally, a mini mental state examination (MMSE) score greater than 24; (4) relatively preserved daily living activities as measured by the Lawton scale; (5) not sufficiently impaired, cognitively and functionally to meet criteria for dementia. Age and years of education were matched to the SMC group. According to their clinical and neuropsychological profile, all patients in this group were considered multi-domain MCI patients (see \cite{pet2004}). As for the geriatric depression scale, none of the MCI showed depression (score lower than 9) \cite{yes1991}.

In the Sternberg's letter-probe task \cite{det1991,mae2001} a set of five letters was presented and participants were asked to keep the letters in mind. A series of single letters ($1 s$ in duration with a random ISI between $2$ and $3$ $s$) was then introduced one at a time, and participants were asked to press a button with their right hand when a member of the previous set was detected. 

Thereafter, single trial epochs were visually inspected by an experienced investigator, and epochs containing visible blinks, eye movements or muscular artifacts were excluded from further analysis. Thirty-five epochs corresponding to each subject were used in order to calculate the functional connectivity values (i.e., their synchronization likelihood). This lower bound was determined by the participant with least epochs. To have an equal number of epochs across participants, thirty-five epochs were randomly chosen from each of the other participants.
The effect of plasticity in the evaluation of network synchronization is not dealt with in this work, although it could be a subject of analysis in future ones. 

Before the MEG recordings, all participants or legal representatives gave written consent to participate in the study, which was approved by the local ethics committee of the Hospital Cl\'inico San Carlos (Madrid, Spain).


\section{Measuring node centrality}
\label{centrality}

The emergence and evolution of MCI has been studied extensively during the last years \cite{pet2004,dic2005,bab2006,sch2007,sch2009,baj2010} since it is known to be related to prodromal Alzheimer's disease (AD).
Neuropathological studies indicate that MCI patients have clear pathophysiological characteristics,
such as the presence of neurofibrillary tangles, loss of dendritic
spines and the accumulation of beta-amyloid protein in the
associative cortex \cite{mar2010}. From the point of view of brain connectivity, MEG recordings of patients
suffering from MCI revealed
an enhancement of the synchronization between cortical regions when memory tasks were
performed \cite{baj2010}. The increase of the synchronized behaviour was accompanied with a change
of the topological structure of the associated functional network, which turned into more random configurations 
\cite{bul2011}. A reduction of the modular behaviour of the network together with an increase
of the long-range functional connections has also been associated with the appearance of MCI \cite{bul2011}.

Nevertheless, how functional hubs are affected by the disease is still unclear. It is known
that network hubs are strongly affected by brain diseases like schizophrenia \cite{bas2008} or 
Alzheimer's disease \cite{buc2009,sta2009,deh2012}. This is bad news, since
the targeted attack to leading nodes leads to a fast damage of the whole network properties \cite{alb2000} and, eventually, to cascading failures \cite{mot2002}.

We used four different centrality measures from complex weighted-network theory
to detect network hubs and to evaluate how their topological roles are affected by the disease. These measures were computed over the matrices
obtained from the patients and control individuals.

\begin{figure}[tp]
\begin{center}
\vskip1.0cm
\psfig{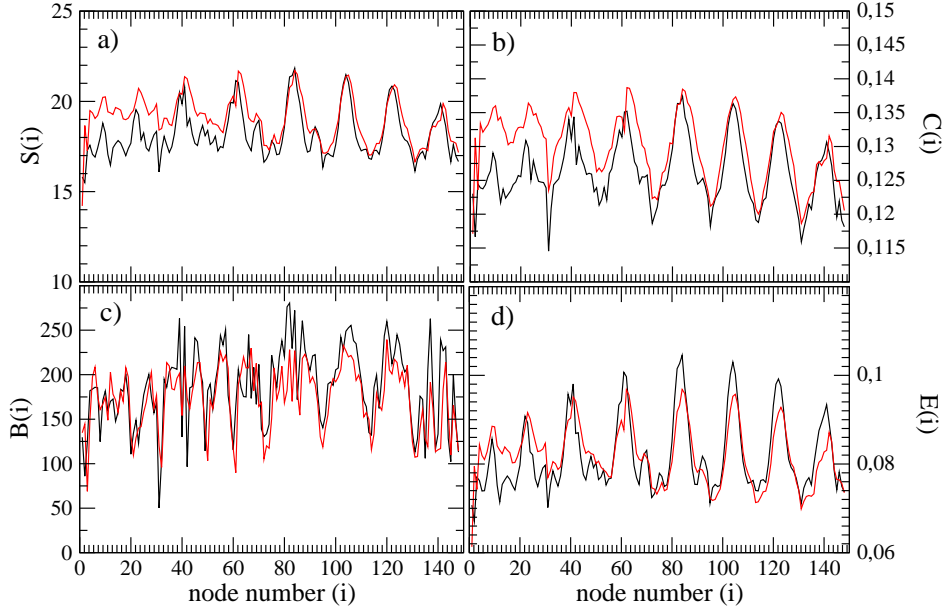}
\vskip0.5cm
\end{center}
\caption{Averaged centrality measures for the control (black) and MCI (red) groups. Specifically, we calculate the node strength
$S(i)$ (a), node closeness $C(i)$ (b), node betweenness $B(i)$ (c) and node eigenvector centrality $E(i)$ (d). All measures have been calculated and averaged over each epoch and for each subject of both groups. Note that $E(i)$ is the measure that better captures the differences between groups in those nodes with higher centrality. See Fig. \ref{fig:fig02} to locate the position of the hubs. 
The periodicity reported in the four figures is only a consequence of the node numbering. See Fig. 2 for details on where nodes are placed in the Euclidean space.
}
\label{fig:fig03}
\end{figure}

The most straightforward centrality measure is the strength of nodes, $S(i)$, defined as the sum of the weights
of the links to all neighbors of a node $i$. If a node has strong connections with its neighbors, it will have higher influence on the functional network, i.e. it will be more central. $S(i)$ can be easily obtained as:
\begin{equation}
S(i)=\sum_{j\in N(i)}W_{ij}
\end{equation}
where $N(i)$ is the set of neighbors of $i$, in our case, all nodes of the network, since ${\bf W}$ matrices are fully connected. Figure \ref{fig:fig03}A shows the strength $S(i)$ of both control and MCI groups, averaged
over the whole groups. Network hubs can be easily identified as those with higher network strength, while the influence of the disease in the hubs is directly the difference of $S(i)$ between both 
groups. We can observe that strength of the network hubs (highest peaks of the $S(i)$ distribution) is not specially altered, while from node 1 to 40 (localized at the central lobe) there is a significant increase of
strength, although they are not hubs of the network. Therefore, node strength seems to indicate that MCI is not specially severe with the 
hubs.

Nevertheless, network strength $S(i)$ is a local measure and
does not accounts for correlations of link strength or specific structural distributions of the weights in the network.
To overcome this issue, two measures of centrality have been applied, with the aim of including 
the influence of the network structure from a point of view of information
transfer: the node closeness $C(i)$ and the node betweenness
$B(i)$. Both measures are based on the geodesic distance from node to node $d_{ij}$, which is obtained
as the shortest distance to arrive from one node $i$ to any other ($j$) by jumping through nodes of the network. At each jump, the traveled distance is increased by the inverse of the weight of the link that is used to pass from one node to the other (e.g., when jumping from node $k$ to node $l$ the travelled distance increases $\Delta d (kl) = 1/w_{kl}$). Finally, the mean geodesic distance of a node $i$, is defined as $l_i=\frac{1}{n}\sum_{j}d_{ij}$. If the mean of geodesic paths of vertex $i$ is low, it means that this vertex can interchange information faster and farther within the network. Closeness centrality $C(i)$ is then defined as the inverse of the mean geodesic path, 
\begin{equation}
C(i)=\frac{1}{l_i}=\dfrac{n}{\sum_{j}d_{ij}}
\end{equation}

A closely related, but not equal, measure of node importance is the betweenness centrality $B(i)$, which measures the ability of a node to  act as transmitter of information through the network. $B(i)$ is defined as  
\begin{equation}
B(i)=\sum_{j\ne k\ne i}\frac{n_{jk}^{i}}{g_{jk}}
\end{equation}

where $n_{jk}^{i}$ accounts for the number of shortest paths between every pair of nodes $j$ and $k$ that pass through node $i$, properly normalized by the total number of geodesic paths $g_{jk}$ from $j$ to $k$. 
 
In Fig. \ref{fig:fig03}B-C we plot the average values of $C(i)$ and $B(i)$ for control and MCI groups. While both measures have their proper characteristics, a common behavior appears in the high correlation present among the centralities of control and MCI for all nodes. The position of the network hubs perfectly matches when
comparing the local measure of centrality $S(i)$ with $C(i)$, while $B(i)$ seems to suggest the existence
of hubs but with a more noisy distribution. Interestingly, the peaks reported in the three distributions correspond to nodes placed at the occipital lobe (see node number in Fig. \ref{fig:fig02}). Again, there are no
significant differences in the centrality of the network hubs, and the most clear signature of MCI is an 
increase of centrality in the nodes belonging to the central lobe (from 1 to 40). This alteration is not
captured by $B(i)$, thus indicating that it is not the most suitable measure to evaluate changes in network centrality.

\begin{figure}[tp]
\begin{center}
\vskip1.0cm
\psfig{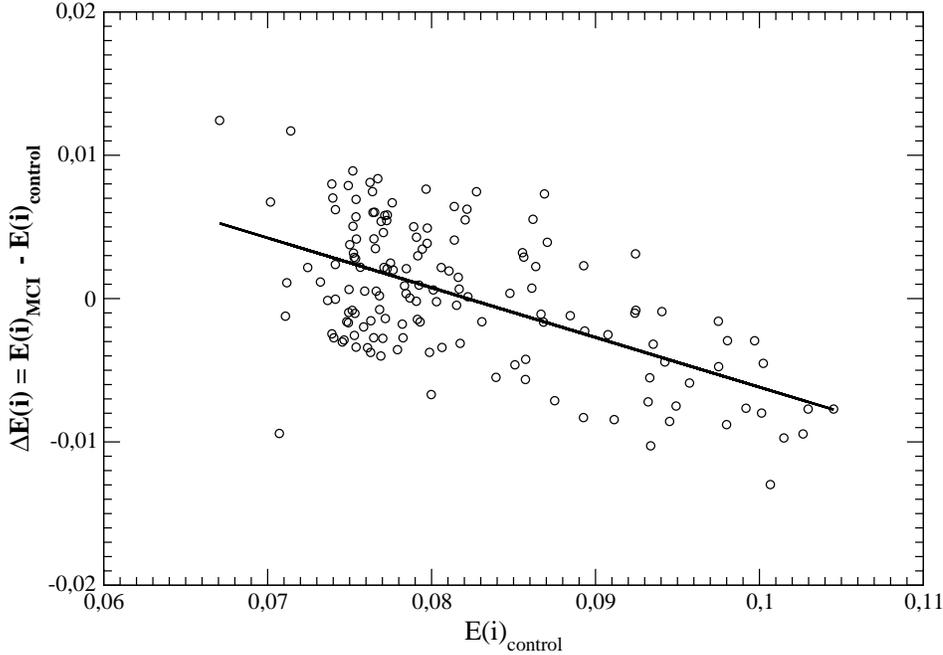}
\vskip0.5cm
\end{center}
\caption{Correlation between the eigenvector centrality $E(i)$ and its variation $\Delta E(i)=E(i)_{MCI}-E(i)_{control}$. Note the
negative correlation indicating that nodes with higher eigenvector centrality are those that are affected the most by the disease.
}
\label{fig:fig04}
\end{figure}

It is worth noting that although $C(i)$ and $B(i)$ contain information about how weights are distributed within the network, they only refer to shortest paths distribution and disregard other structural properties of the network.
To overcome this issue, we compute the eigenvector centrality $E(i)$ of the nodes. $E(i)$ is a measure
obtained from the spectral analysis of the connectivity matrix ${\bf W}$. Specifically, it is calculated as the eigenvector $v_1$ of the first eigenvalue of the matrix ${\bf W}$. Note that any modification of the weight of the links, will result in a change of the matrix ${\bf W}$ and, therefore, will be reflected in the value
of the eigenvector centrality $E(i)$. On the contrary, if an alteration of a link does not imply any modification in the distribution of shortest paths, it will not be captured by $C(i)$ or $B(i)$ centralities.
The mathematical definition of the eigenvector centrality 
is related to an iterative process were the centrality of a node $i$ is calculated as the sum of the centralities of its neighbors: 
\begin{equation}
 E(i)= \lambda^{-1} \sum_j W_{ij} E(j),
\label{eq:centrality}
\end{equation}
where $\lambda$ is a constant, $E(j)$ is the eigenvector centrality of node $j$, being $j$ a neighbour of $i$ and $W_{ij}$ are the components of the connectivity matrix ${\bf W}$. In matrix notation, Eq. \ref{eq:centrality} reads
{\bf $\lambda E= W E$} so that {\bf $E$} can be expressed as a linear combination of the eigenvectors {\bf $v_i$} of the adjacency
matrix. Since the final value of $E(i)$ is obtained when $t \rightarrow \infty$, the value of $E(\infty)$ is equal to the eigenvector $v_1$ associated to the dominant eigenvalue $\lambda_1$. Therefore, the eigenvector centrality $E(i)$ is equal to the eigenvector $v_1(i)$ of the connectivity matrix $W$ \cite{new2004}. 

\begin{figure}[tbp]
\begin{center}
\psfig{file=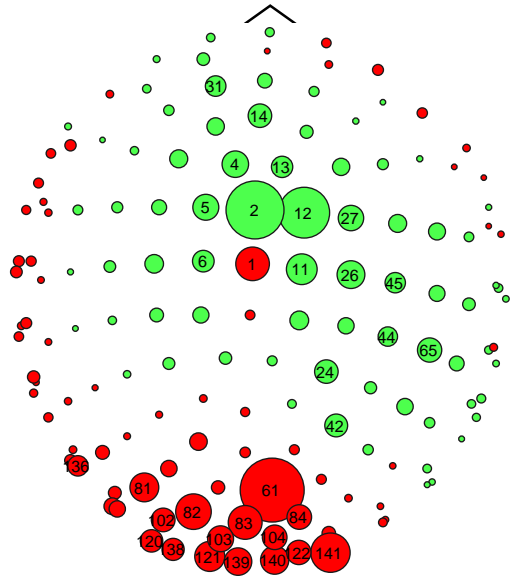,width=0.4\textwidth}
\end{center}
\caption{Distribution of the variation of eigenvector centrality $E(i)$. The size of the nodes is proportional 
to $|\Delta E(i)|=|E(i)_{MCI}-E(i)_{control}|$, while node color indicates an increase (green) or decrease (red) in the node
centrality. Node number is indicated for those nodes with high $|\Delta E(i)|$.}
\label{fig:fig05}
\end{figure}

Figure \ref{fig:fig03}D shows the eigenvector centrality $E(i)$ for both groups. It detects the position of the network hubs which, as in $S(i)$ and $C(i)$, are mainly placed at the occipital lobe. Interestingly, this measure captures a clear decay in the centrality of the hubs, as indicated by the decrease of the peaks height. The consequence of this decrease is a network were the hubs play a less relevant role. In Fig. \ref{fig:fig04} we 
plot how the variation of the eigenvector centrality $\Delta E (i)$ due to MCI is related with the initial centrality of the nodes. We can observe a negative correlation, indicating that those nodes with higher centrality are, in turn, the nodes that decrease its importance the most. It is also interesting to see what is the position of the nodes in the whole network. Figure \ref{fig:fig05} shows the increase (green) and decrease
(red) of eigenvector centrality with the node size proportional to the value of $|\Delta E(i)|$. It is the occipital lobe the one that is mostly affected by the decrease of centrality, while the central lobe takes
advantage of this reduction.

\section{Conclusions}
\label{discussion}

The analysis of functional and anatomical brain networks using complex networks analysis have revealed
interesting information about how these networks are organized \cite{bul2009}. All studies agree on the fact that
brain networks are highly heterogeneous, giving rise to the existence of hubs, i.e. leading nodes
from the point of view of the network structure. In this manuscript we have shown a detailed analysis of how
the importance of network hubs in a functional network associated with the execution of a memory task
is affected by the
emergence of MCI. We report how the disease particularly affects network hubs, reducing their importance
in the network. 
This reduction is captured by all different measures of centrality, independently of whether they rely on the local or global properties of the network.
Among all centrality measures, eigenvector centrality $E(i)$ is the one that better captures
the effect of the disease on the network hubs. 
The occipital region, containing the majority of the hubs in the case of healthy individuals, 
is the region where centrality decreased the most, while nodes belonging to the central lobe benefit from the hub deterioration. Interestingly, anatomical networks of patients suffering from schizophrenia also show 
a reduction of the hub importance, together with a dispersion in the location of network hubs \cite{bas2008}.
This reduction of hub centrality can be used as a signature of the existence of pathology since it is known
that the hub structure is quite stable in healthy adult individuals \cite{hwa2012,zuo2012}. 
In addition, it reinforces the hypothesis that MCI increases the randomness and 
homogeneity of the functional network \cite{bul2011}, since the reduction of hub importance leads to more homogeneous networks.
The implications of hub deterioration are still unclear. For example, in Alzheimer's disease, it has been shown that amyloid-beta deposition in the locations of cortical hubs could act critically in the severity of the disease
\cite{buc2009}. In the case of anatomical networks in schizophrenia, the reduction of centrality of the 
frontal hubs has been related to the disorganization of the anatomical network.
The reduction in hub centrality in posterior scalp regions could be related to the high levels of beta amyloid accumulation in those regions found in MCI and Alzheimer's disease.
Conversely, the increased hub centrality found for anterior sensors could reflect enhanced engagement of frontal regions compensating for decreased capacity to tackle the demands of the memory task (see \cite{baj2010} for a similar interpretation).
We believe that our work will shed some light on the 
role played by hubs in brain networks, which may have strong influence on the network 
robustness \cite{alb2000}, cascading processes \cite{mot2002} and network controllability \cite{tan2012}.
Similar studies could be carried out in other neurodegenerative diseases where the existence of functional hubs has been reported.





\nonumsection{Acknowledgments} 
This work was supported by the Spanish Ministry of S\&T [FIS2009-07072], the Community of Madrid 
under the R\&D Program of activities MODELICO-CM [S2009ESP-1691], the Spanish Ministry of economy and competitiveness 
[PSI2012-38375-C03-01], as well to a grant from the CAM [S2010/BMD-2460].


\newpage

\begin{thebibliography}{9}


\bibitem[Achard {\it et al.} (2006)]{ach2006}
Achard, S., Salvador, R., Whitcher, B., Suckling, J. \& Bullmore, E. [2006]
``A Resilient, Low-Frequency, Small-World Human Brain Functional Network with Highly Connected Association Cortical Hubs'', 
{\it Journal of Neuroscience} {\bf 26}, pp. 63--72.

\bibitem[Albert {\it et al.} (2000)]{alb2000}
Albert, R., Jeong, H. \& Barab\'asi, A.-L. [2000]
``Error and attack tolerance of complex networks'',
{\it Nature} {\bf 406}, pp. 378-382.

\bibitem[Albert \& Barab\'asi (2007)]{alb2007}
Albert, R. \& Barab\'asi, A.-L. [2007]
``Statistical mechanics of complex networks'',
{\it Rev. of Mod. Phys.} {\bf 74}, pp. 47-97.

\bibitem[Babiloni {\it et al.} (2006)]{bab2006}
Babiloni, C., Ferri, R., Binetti, G., Cassarino, A., Dal Forno, G., et al. [2006] 
``Frontoparietal coupling of brain rhythms in mild cognitive impairment: A multicentric
eeg study'', 
{\it Brain Res. Bull.} {\bf 69}, pp. 63--73.

\bibitem[Bajo {\it et al.} (2010)]{baj2010}
Bajo, R., Maest\'u, F., Nevado, A., Sancho, M., Guti\'errez, R., et al. [2010] 
``Functional connectivity in mild cognitive impairment during a memory task: implications
for the disconnection hypothesis'', 
{\it J. Alzheimer Dis.} {\bf 22}, pp. 183--93.

\bibitem[Barab\'asi \& Albert (1999)]{bar1999}
Barab\'asi, A.-L. \& Albert, R. [1999] 
``Emergence of scaling in random networks'', 
{\it Science} {\bf 286}, pp. 509--572.

\bibitem[Basset {\it et al.} (2008)]{bas2008}
Bassett, D.S., Bullmore, E., Verchinski, B.A., Mattay, V.S., Weinberger, D.R. \& Meyer-Lindenberg, A. [2008]
``Hierarchical organization of human cortical networks in health and schizophrenia'',
{\it J Neurosci.} {\bf 28},pp. 9239--48.

\bibitem[Boccaletti {\it et al.} (2002)]{boc2002}
Boccaletti, S., Kurths, J., Osipov, G., Valladares, D.L. \& Zhou, C.S. [2002]
``The synchronization of chaotic systems'',
{\it Phys. Rep.} {\bf 366}, pp. 1-101.

\bibitem[Boccaletti {\it et al.} (2006)]{boc2006}
Boccaletti, S., Latora, V., Moreno, Y., Chavez, M. \& Hwang, D.-U. [2006] 
``Complex networks: Structure and dynamics'', 
{\it Phys. Rep.} {\bf 424}, pp. 175--308.

\bibitem[Breakspear \& Terry (2009)]{bre2002}
Breakspear, M. \& Terry, J.R. [2002]
``Nonlinear interdependence in neural systems: motivation, theory, and relevance'',
{\it Int. J. of Neurosci.} {\bf 112}, pp. 1263-1284.

\bibitem[Buckner {\it et al.} (2009)]{buc2009}
Buckner, R.L., Sepulcre, J., Talukdar, T., Krienen, F.M., Liu, H., Hedden, T., Andrews-Hanna, J.R., Sperling, R.A. \& Johnson, K.A. [2009]
``Cortical hubs revealed by intrinsic functional connectivity: mapping, assessment of stability, and relation to Alzheimer's disease'',
{\it J Neurosci.} {\bf 29}, pp. 1860--73.

\bibitem[Bullmore \& Sporns (2009)]{bul2009}
Bullmore, E. \& Sporns, O. [2009]
``Complex brain networks: graph theoretical analysis of structural and
functional systems'', 
{\it Nature Reviews} {\bf 10}, 186.

\bibitem[Buld\'u {\it et al.} (2011)]{bul2011}
Buld\'u, J.M., Bajo, R., Maest\'u, F., Castellanos, N., Leyva, I., Gil, P., Sendi\~na-Nadal, I., Almendral, J.A., Nevado, A., del-Pozo, F. \& Boccaletti, S. [2011]
``Reorganization of Functional Networks in Mild Cognitive Impairment'',  
{\it PLoS ONE} {\bf 6}, e19584.

\bibitem[Buzug {\it et al.} (1994)]{buz1994}
Buzug, Th., Pawelzik, K., von Stamm, J. \& Pfister, G. [1994]
``Mutual information and global strange attractors in Taylor-Couette flow'',
{\it Physica D} {\bf 72}, pp. 343-–350.

\bibitem[de Haan {\it et al.} (2012)]{deh2012}
de Haan, W., Mott, K., van Straaten, E.C., Scheltens, P. \& Stam, C.J. [2012] 
``Activity dependent degeneration explains hub vulnerability in Alzheimer's disease'',
{\it PLoS Comput. Biol.} {\bf 8}, e1002582.

\bibitem[deToledo-Morrell {\it et al.} (1991)]{det1991}
deToledo-Morrell, L., Evers, S., Hoeppner, T.J., Morrell, F., Garron, D.C. \& Fox, J.H. [1991]
``A 'stress' test for memory dysfunction. Electrophysiologic manifestations of early Alzheimer's disease'',  
{\it Arch. Neurol.} {\bf 48}, p. 605–-609.

\bibitem[Dickerson {\it et al.} (2005)]{dic2005}
Dickerson, B.C., Salat, D.H., Greve, D.N., Chua, E.F., Rand-Giovannetti E., et al. [2005]
``Increased hippocampal activation in mild cognitive impairment
compared to normal aging and AD'', 
{\it Neurology} {\bf 65}, pp. 404--411.

\bibitem[Grundman {\it et al.} (2004)]{gru2004}
Grundman, M., Petersen, R.C., Ferris, S.H., Thomas, R.G., Aisen, P.S., Bennett, D.A., et al [2004]
``Mild cognitive impairment can be distinguished from Alzheimer disease and normal aging for clinical trials'',  
{\it Arch. Neurol.} {\bf 61}, p. 59–-66.

\bibitem[Guimer\`a \& Amaral (2005)]{gui2005}
Guimer\`a, R., \& Amaral, L.A.N. [2005] 
``Functional Cartography of complex metabolic networks'', 
{\it Nature} {\bf 433}, p. 895.

\bibitem[Hwang {\it et al.} (2012)]{hwa2012}
Hwang, K., Hallquist, M.N. \& Luna, B. [2012]
``The Development of Hub Architecture in the Human Functional Brain Network'',
{\it Cereb Cortex.} published on-line 2012 Aug 28. doi:10.1093/cercor/bhs227. 

\bibitem[Joyce {\it et al.} (2010)]{joy2010}
Joyce, K.E., Laurienti, P.J., Burdette, J.H. \& Hayasaka, S. [2010] 
``A New Measure of Centrality for Brain Networks'', 
{\it PLoS ONE} {\bf 5}, e12200.

\bibitem[Kuhnert {\it et al.} (2012)]{kuh2012}
Kuhnert, M.T., Geier, C., Elger, C.E. \& Lehnertz, K. [2012]
``Identifying important nodes in weighted functional brain networks: a comparison of different centrality approaches'',
{\it Chaos} {\bf 22}, p. 023142.

\bibitem[Lohmann {\it et al.} (2010)]{loh2010}
Lohmann, G., Margulies, D.S., Horstmann, A., Pleger, B., Lepsien, J. et al. [2010] 
``Eigenvector Centrality Mapping for Analyzing Connectivity Patterns in fMRI Data of the Human Brain'',
{\it PLoS ONE} {\bf 5}, e10232.

\bibitem[Markesbery (2010)]{mar2010}
Markesbery, W. [2010] 
``Neuropathologic alterations in mild cognitive impairment: A review'',
{\it J. Alzheimer Dis.} {\bf 19}, pp. 221--228.

\bibitem[Maestu {\it et al.} (2001)]{mae2001}
Maestu, F., Fernandez, A., Simos, P.G., Gil-Gregorio, P., Amo, C., Rodriguez, R., Arrazola, J. \& Ortiz, T. [2001]
``Spatio-temporal patterns of brain magnetic activity during a memory task in Alzheimer's disease'',  
{\it Neuroreport} {\bf 12}, p. 3917–-3922.

\bibitem[Motter \& Lai	 (2002)]{mot2002}
Motter, A.E. \& Lai, Y.C. [2002]
``Cascade-based attacks on complex networks'',
{\it Phys. Rev. E} {\bf 66}, p. 065102.

\bibitem[Newman (2003)]{new2003}
Newman, M.E.J. [2003]
``The structure and function of complex networks'', 
{\it SIAM Rev.} {\bf 45}, pp. 67--256.

\bibitem[Newman (2004)]{new2004}
Newman, M.E.J. [2004]
``Analysis of weighted networks'',
{\it Phys. Rev. E} {\bf 70}, p. 056131.

\bibitem[Pereda {\it et al.} (2005)]{per2005}
Pereda E, Quian Quiroga R, Bhattacharya B [2005] 
``Nonlinear multivariate analysis of neurophysiological signals'', 
{\it Progress in Neurobiology} {\bf 77}, pp. 1--37. 

\bibitem[Petersen {\it et al.} (2004)]{pet2004}
Petersen R.C. [2004]
``Mild cognitive impairment as a diagnostic entity'',  
{\it J. Intern. Med.} {\bf 256}, p. 183-–194.

\bibitem[Scheff {\it et al.} (2007)]{sch2007}
Scheff, S.W., Price, D.A., Schmitt, F.A., DeKosky, S.T. \& Mufson, E.J. [2007]
``Synaptic alterations in ca1 in mild alzheimer disease and mild cognitive impairment'',
{\it Neurology} {\bf 68}, pp. 1501--1508.

\bibitem[Scheneider {\it et al.} (2009)]{sch2009}
Schneider, J.A., Arvanitakis, Z., Leurgans, S.E., Bennett, D.A. [2009]
``The neuropathology of probable alzheimer disease and mild cognitive impairment'',
{\it Ann. Neurol.} {\bf 66}, pp. 200--208.

\bibitem[Sporns {\it et al.} (2007)]{spo2007}
Sporns, O., Honey, Ch.J. \& Kotter, R. [2007] 
``Identification and classification of hubs in brain networks'',
{\it PLoS ONE} {\bf 2}, p. e1049.

\bibitem[Stam \& van Dijk (2002)]{sta2002}
Stam C. \& van Dijk, B.W. [2002] 
``Synchronization likelihood: an unbiased measure of generalized synchronization in multivariate data sets'', 
{\it Physica D} {\bf 163}, pp. 236–-251.

\bibitem[Stam {\it et al.} (2006)]{sta2006}
Stam, C.F., Jones, B.F., Manshanden, I., van Cappellen van Walsum, A.M., Montez, T., Verbunt, J.P.A., de Munck, J.C., van Dijk, B.W., Berendse, H.W. \& Scheltens, P. [2006] 
``Magnetoencephalographic evaluation of resting-state functional connectivity in Alzheimer's disease'',
{\it NeuroImage} {\bf 32}, pp. 1335-1344.

\bibitem[Stam {\it et al.} (2009)]{sta2009}
Stam, C.J., de Haan, W., Daffertshofer, A., Jones, B.F., Manshanden, I., van Cappellen van Walsum, A.M., Montez, T., Verbunt, J.P., de Munck, J.C., van Dijk, B.W., Berendse, H.W. \& Scheltens, P. [2009] 
``Graph theoretical analysis of magnetoencephalographic functional connectivity in Alzheimer's disease'', 
{\it Brain} {\bf 132}, pp. 213--224. 

\bibitem[Tang {\it et al.} (2012)]{tan2012}
Tang, Y., Gao, H., Zou, W. \& Kurths, J. [2012] 
``Identifying Controlling Nodes in Neuronal Networks in Different Scales'', 
{\it PLoS ONE} {\bf 7}, e41375.


\bibitem[Watts \& Strogatz (1998)]{wat1998}
Watts, D.J. \& Strogatz, S.H. [1998] 
``Collective dynamics of small-world networks'', 
{\it Nature} {\bf 393}, pp. 440--442.

\bibitem[Yesavage \& Brooks (1991)]{yes1991}
Yesavage, J.A. \& Brooks, J.O. [1991]
``On the importance of longitudinal research in Alzheimer's disease'',  
{\it J Am Geriatr Soc} {\bf 39(9)}, p. 942-–944.

\bibitem[Zamora-L\'opez {\it et al.} (20010)]{zam2010}
Zamora-L\'opez, G., Zhou, C. \& Kurths, J. [2010]
``Cortical hubs form a module for multisensory integration on top of the hierarchy of cortical networks'',
{\it Front. Neuroinformatics} {\bf 4}, 1.

\bibitem[Zuo {\it et al.} (2012)]{zuo2012}
Zuo, X.N., Ehmke, R., Mennes, M., Imperati, D., Castellanos, F.X., Sporns, O. \& Milham, M.P. [2012]
``Network centrality in the human functional connectome'',
{\it Cereb Cortex.} {\bf 22} pp. 1862--75.

\end{thebibliography}
\end{document}